\documentclass[runningheads]{llncs}
\usepackage{graphicx}
\usepackage{xcolor}
\usepackage{amsmath}
\begin{document}
\title{MVBIND: SELF-SUPERVISED MUSIC RECOMMENDATION FOR VIDEOS VIA EMBEDDING SPACE BINDING}
%
%
\author{Jiajie Teng\inst{1} \and
Huiyu Duan\inst{1} \and
Yucheng Zhu\inst{1} \and
Sijing Wu\inst{1} \and
Guangtao Zhai\inst{1}}
\authorrunning{Jiajie Teng et al.}
\institute{Shanghai Jiao Tong University, Shanghai, China \\
\email{1584012901, huiyuduan, zyc420, wusijing, zhaiguangtao@sjtu.edu.cn}}

\maketitle   
\begin{abstract}
Recent years have witnessed the rapid development of short videos, which usually contain both visual and audio modalities.
Background music is important to the short videos, which can significantly influence the emotions of the viewers. However, at present, the background music of short videos is generally chosen by the video producer, and there is a lack of automatic music recommendation methods for short videos. This paper introduces \textbf{MVBind}, an innovative \textbf{\underline{M}}usic-\textbf{\underline{V}}ideo embedding space \textbf{\underline{Bind}}ing model for cross-modal retrieval. MVBind operates as a self-supervised approach, acquiring inherent knowledge of intermodal relationships directly from data,  without the need of manual annotations. Additionally, to compensate the lack of a corresponding musical-visual pair dataset for short videos, we construct a dataset, SVM-10K(Short Video with Music-10K), which mainly consists of meticulously selected short videos. On this dataset, MVBind manifests significantly improved performance compared to other baseline methods. The constructed dataset and code will be released to facilitate future research.

\keywords{MVBind \and self supervised pre-training \and  music recommendation \and  cross modal retrieval \and  short videos.}
\end{abstract}
\section{Introduction}
\label{sec:intro}

With the advancement of Internet communication, short videos have gradually become the most popular video products.
Music is an essential component in short videos, which significantly influences the emotion, understanding, and experience of the viewers.
Generally, the background music of short videos is selected by video producers, which consumes a lot of their time and energy.
However, with the increasing of music and video data, video creators are confronted with the difficulty of selecting appropriate music for different segments of their videos.
Therefore, a music recommendation system for short videos is crucial for improving the efficiency of producing short videos.

Classical music recommendation methods generally adopt single-modal retrieval methodologies, which employ approaches such as keyword-based and content-based searching~\cite{zenggang2021research}. These methods are limited to assessing similarity within discrete groups, such as image~\cite{garg2021novel}, text~\cite{luan2021sparse}, video~\cite{dzabraev2021mdmmt}, and audio~\cite{castellon2021codified} searches. Furthermore, even when the user possesses a clear
understanding of the desired music genre, utilizing conventional text-based methodologies for music selection is still complicated and time-consuming, since it is difficult to completely label the emotion of a song through a few words and appropriately retrieve the music through the texts.
Therefore, a musical-visual alignment model is desired, which can better promote the performance of the music recommendation system.

Currently, most cross-modal audio-visual pre-training works mainly focus on establishing a semantic correspondence between visual and auditory modalities~\cite{girdhar2023imagebind} (such as matching a dog image with the sound of a dog barking). However, the problem of music recommendation for short videos discussed
in this article is more complex than a simple semantic correspondence~\cite{alayrac2020self,alwassel2020self}, since it involves higher-level connections, such as emotions, overall visual style, aesthetics, rhythm, \textit{etc}. Moreover, it is also important to provide suitable but diverse background music based on a given video, which can inspire more creative ideas and possibilities for short video creators. 

To address the lack of research in this area, in this paper, we propose \textbf{MVBind}, a \textbf{\underline{M}}usic-\textbf{\underline{V}}ideo self-supervised embedding space \textbf{\underline{Bind}}ing model, to achieve cross-modal retrieval between the music and video modalities.
To fill the absence of large-scale short video datasets for cross-modal content retrieval, we first construct a short video with music-10K (SVM-10K) dataset, which contains nearly 10k short videos with more than 10k stars or likes.
A strict screening procedure and a pre-processing method are conducted for the dataset to ensure the quality and correspondence of the music and videos.
Based on the established dataset, we propose a self-supervised music and video embedding space connection method, which first extracts features from ImageBind model and uses contrastive learning to bind two spaces.
We also conduct a comprehensive experiment to explore whether and how different pre-training models influence the cross-modal retrieval performance.
Our main contributions are summarized as follows:
\begin{itemize}

  \item We establish a large-scale short video with music (SVM-10K) dataset, which has nearly 10k short videos with more than 10k stars.

  \item A self-supervised music-video cross-modal binding model termed MVBind is proposed for cross-modal retrieval.

  \item We conduct comprehensive experiments using existing unimodal video and audio neural networks to explore the combination performance of different feature extraction methods.
\end{itemize}

\section{Related Work}
\subsection{Audio-visual Video Dataset}
Classical video datasets mainly contain videos with real captured visual and audio content, which have semantic-level connections. For instance, Kinetics400~\cite{carreira2017quo} contains diverse action scenes with audio tracks. The YouTube-8M dataset~\cite{abu2016youtube} comprises 8 million video segments collected from YouTube.
However, these datasets are proposed for general video tasks, and their audio track styles in the videos are excursive, while our task focuses on the background music recommendation.
Meanwhile. Some works have also constructed datasets for audio-video cross-modal retrieval. For instance, Dídac Surís \textit{et al.}~\cite{suris2022s}have constructed an audio-video retrieval dataset by collecting data from two datasets, \textit{i.e.}, YT8M-MusicVideo and MovieClips.However, the video clips they extracted are excessively brief (often spanning only a few hundred milliseconds to a few seconds). Additionally, the video-audio correlations in the dataset lack considerations for mood or rhythmic styles and overlook constraints such as the number of likes/favorites of the original video, thereby failing to adequately represent popular aesthetics.
Daniel McKee \textit{et al.}~\cite{mckee2023language} have established a dataset with 4k videos and corresponding texts extracted from YT8M-MusicVideo. Overall, all these datasets are collected from general video datasets, and they lack a large-scale high-quality music-video cross-modal retrieval dataset for short videos, which mainly focuses on higher-level emotional-level connections.

\subsection{Audio-Visual Crossmodal Studies}
\textbf{Audio feature Extraction Network.} In the realm of audio processing, researchers have introduced diverse and robust network structures, laying a solid foundation for tasks such as audio analysis, classification, and generation, \textit{etc}. The classic Mel-Frequency Cepstral Coefficients (MFCC) method, proposed by Muda \textit{et al.}~\cite{muda2010voice}, is widely employed for extracting spectral features from speech signals. In 2017, the Google team introduced VGGish, a network that utilizes a convolutional structure to transform audio segments into 128-dimensional embedding vectors, capturing advanced semantic information.
Baevski \textit{et al.} have presented wav2vec2.0~\cite{baevski2020wav2vec}, a model specifically designed for self-supervised learning of audio representations.

\setlength{\parindent}{0pt}\textbf{Visual feature Extraction Network}. Visual Recognition has achieved significant advancements in recent years, driven by the evolution of deep neural networks (DNNs) and human vision studies~\cite{min2024perceptual,duan2024quick,duan2023attentive,duan2022develop,duan2019visual,tu2022end,tu2022iwin,fang2020identifying}.
Many works have been proposed to improve the architecture of DNNs, such as ResNet~\cite{he2016deep}, EfficientNet~\cite{tan2019efficientnet}, Vision Transformer~\cite{dosovitskiy2020image}, \textit{etc.}
Moreover, some works have also studied the influence of different pre-training methods on the performance of DNNs, such as imagenet pre-training~\cite{duan2022confusing}, vector-quantized pre-training~\cite{duan2022saliency}, contrastive learning~\cite{he2020momentum}, mask pre-training~\cite{he2022masked,duan2023masked}, \textit{etc.}

\setlength{\parindent}{0pt}\textbf{Music Recommendation Work}.
Although cross-modal research has been widely studied and applied~\cite{sun2024visual,sun2023influence,yang2024aigcoiqa2024,wang2023aigciqa2023,zhu2023audio,zhu2023perceptual}, there are still few related works on audio-visual cross-modal retrieval.
Sungeun Hong \textit{et al.} have introduced CBVMR~\cite{hong2018cbvmr}, a content-based cross-modal matching method for video and music retrieval using DNNs, which employs inter-modal ranking loss to bring semantically similar videos and music closer in the embedding space. Andrey Guzhov \textit{et al.} have presented AudioCLIP~\cite{guzhov2022audioclip}, a cross-modal audio representation learning model derived from CLIP~\cite{radford2021learning}, replacing the text part with audio.
Zhuo \textit{et al.}~\cite{zhuo2023video} have proposed an audio-video cross-modal learning framework for audio generation rather than retrieval.

\section{SVM-10K Dataset}

To compensate for the lack of corresponding musical-visual pair dataset focus on short videos, we construct SVM-10K dataset (Short Video with Music 10K), which consists of nearly 10000 meticulously selected short videos. With high-quality short videos and large data scales, the proposed SVM-10K dataset fills the absence of large-scale short video datasets for cross-modal content retrieval and further promotes the development of background music retrieval.
In this section, we first introduce how we collect and filter the short videos to build the SVM-10K dataset.

\subsection{Dataset Construction}

\textbf{Data Collection and Filtering.} The primary source of SVM-10K comes from an open social platform, TikTok, which supports search through titles, tags, and other means. Users can browse content without authentication, and specific search terms such as ``travel photography” and ``vlog” can be used to find short videos likely to contain background music. By parsing specific elements of the HTML page~\cite{gojare2015analysis}, the description information for individual short videos can be obtained in bulk. We consider good audiovisual effects as a necessary prerequisite for short videos to receive high likes, thus short videos containing specific keywords and with more than 10,000 likes are selected as potential samples for the SVM-10K dataset. After this step, we collect a total of 40000 short videos.
Furthermore, we manually filtered the collected 40000 short videos for dataset quality. Since this paper focuses on music recommendations, ambient noise is undesirable since it may affect the audio feature extraction process. Simply using voice separation methods is an optional way, however, it may result in the removal of certain sounds of the background music, which will affect the overall audio quality of the dataset. Taking these factors into consideration, we manually filter the collected short videos and finally obtain nearly 10000 videos as our SVM-10K dataset.

\hspace*{2em}The collected short videos generally contain ``video black border" , which refers to the black border that appears in a video, potentially hindering hash value extraction in video retrieval and hashing methods and leading to misjudgments. To address this issue and remove video black border, the following steps are taken.
\textbf{(1)}
\textbf{Calculate Standard Deviation:} Compute the standard deviation of the color histogram. If it falls below a certain threshold, classify the video frame as borderless.
\textbf{(2) Convert to Binary Image:} Utilize the OTSU~\cite{otsu1975threshold} algorithm to convert the image into a binary image, determining foreground and background colors.
\textbf{(3) Sobel Edge Detection:}~\cite{kanopoulos1988design} Apply the Sobel operator for edge detection in horizontal and vertical directions based on the binary image.
\textbf{(4) Edge Analysis:} Analyze the detected edges by calculating the proportion of edge length and average pixel values.
\textbf{(5)Identify Black Border Edge:} Identify the black border edge by folding the edges along the video frame's center point in both directions and checking if the pixel average is not close to a solid color.
\textbf{(6)Non-Maximum Suppression (NMS):} Apply non-maximum suppression (NMS) to obtain a unified border edge from all detected edges within the video.
\textbf{(7)Segment Screenshots:} Use the unified border edge to segment all screenshots, resulting in the final outcome.

\subsection{Data Statistics}
The collected SVM-10K dataset is composed of diverse themes, which are demonstrated as follows.
\textbf{(1)Video Material Suitability:} Videos with a strong correlation between visuals and audio are retained.
\textbf{(2)Sports-related Videos:} Event montages and sports tutorials with significant human voices are excluded.
\textbf{(3)Food-related Videos:} Videos with dominant environmental sounds over background music are excluded, with criteria based on subjective judgment.
\textbf{(4)Relationship-themed Videos:} Videos focusing on couples, pets, or parent-child relationships are retained if they have minimal environmental sounds and lack significant human voices.
\textbf{(5)Ancient Town Scenery Videos:} Most videos meet the criteria, but those with prominent rain/wind sounds related to video content are removed.
Additionally, video titles are retained during crawling for potential future incorporation of text features.

\section{MVBind: Music-video Embedding Space Binding}

\begin{figure*}[t]
    \centering
    \includegraphics[width=\textwidth]{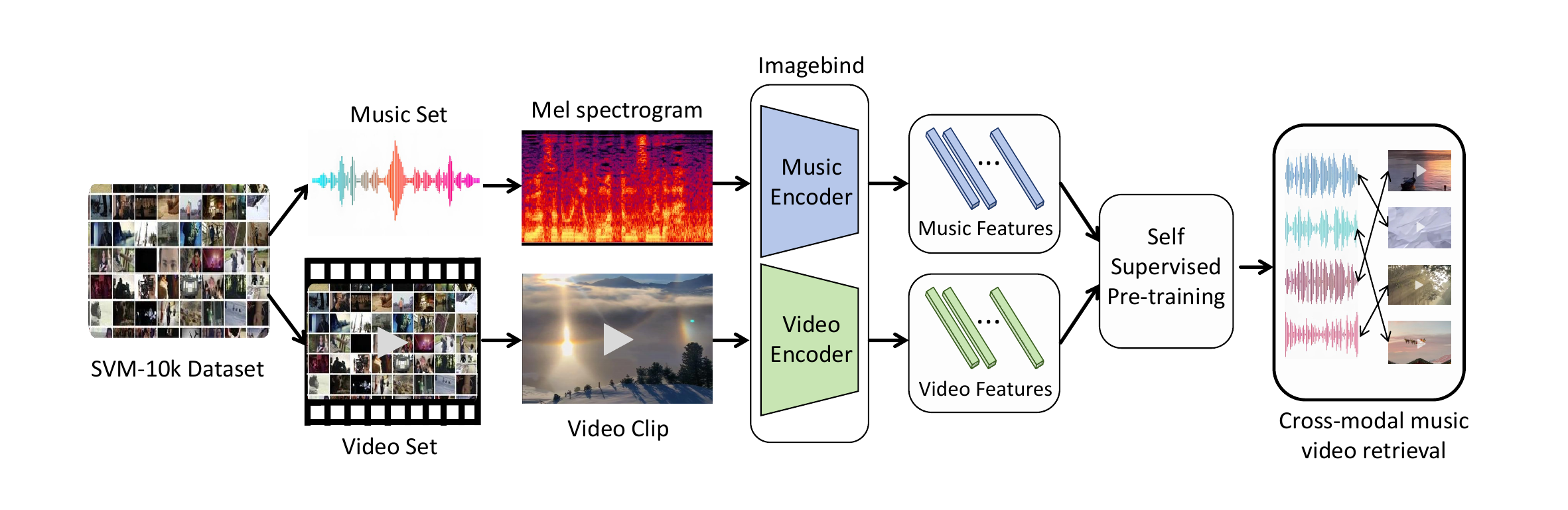}
    \caption{\textbf{An overview of the proposed MVBind.} It first separates the audio and video modalities from short videos in the SVM-10K dataset. For the audio signal, the Mel spectrogram is extracted, and then a 1024-dimensional audio feature is obtained using ViT pre-trained by ImageBind. For the video signal, preprocessing is performed (such as removing black borders), and then a 1024-dimensional video feature is extracted using ViT pre-trained by ImageBind. Self-supervised learning is then used to train and connect the two modal features. The ultimate goal is to achieve cross-modal music video retrieval.}
    \label{fig:figure1}
\end{figure*}

Along with the SVM-10K dataset, we propose a music recommendation algorithm for short videos named MVBind.
In this section, we first introduce the feature extraction operations on the audio set and the preprocessed video set (Section 4.1). Secondly, there is our self-proposed algorithm for self-supervised training of the features extracted from the two modalities (Section 4.2). The overview of MVBind is illustrated in Figure~\ref{fig:figure1}.

\subsection{Multimodal Feature Extraction}
We employ ImageBind~\cite{girdhar2023imagebind}, a comprehensive multimodal binding method that encompasses six modalities, as our feature extractor to extract both the audio and video features. 
While ImageBind primarily focuses on cross-modal classification retrieval, its alignment across six modalities makes it an ideal pre-trained model for feature extraction. Utilizing ImageBind improves the integration of visual and auditory aspects compared to other feature extraction models, which is particularly beneficial for the subsequent self-supervised training of both modalities.

\noindent\textbf{Audio feature extraction.} The audio is encoded using~\cite{gong2021ast}, which transforms a 2-second audio sampled at 16kHz into a spectrogram with 128 mel-frequency bins. Considering the similarity between the spectrogram and the image, a ViT with a patch size of 16 and a stride of 10 is used to extract features for audio.

\noindent\textbf{Video feature extraction.} We apply the Vision Transformer (ViT) for image feature extraction. We utilize 2 frames from a 2-second video clip, which employs the pre-trained ViT model on a large image classification dataset. The 1024-dimensional embedding output from the last layer serves as the feature representation of the video frames in our short video background music retrieval algorithm.
Concretely, ViT partitions the input image into patches (16x16)~\cite{dosovitskiy2020image} and projects each patch into a fixed-length vector. These vectors are then fed into the Transformer~\cite{vaswani2017attention} for further encoding.

\hspace*{2em}Both encoders incorporate modality-specific linear projection heads, yielding fixed-size $d$-dimensional embeddings. These embeddings are normalized and used for the InfoNCE loss :

\begin{equation*}
\mathcal{L}_{\mathcal{I}, \mathcal{M}} = -\log \frac{\exp \left( \mathbf{q}_i^\top \mathbf{k}_i / \tau \right)}{\exp \left( \mathbf{q}_i^\top \mathbf{k}_i / \tau \right) + \sum_{j \ne i} \exp \left( \mathbf{q}_i^\top \mathbf{k}_j / \tau \right)}
\end{equation*}

where $\tau$ is a scalar temperature that controls the smoothness of the softmax distribution and $j$ denotes unrelated observations, also called `negatives'.L and M represent two different modalities.

\hspace*{2em}This setup not only facilitates effective learning but also allows for encoder initialization using other pre-trained models. For instance, one can initialize a subset of encoders using pre-trained models like the image and text encoders of CLIP~\cite{radford2021learning} or OpenCLIP. This design promotes versatility and allows for the utilization of pre-existing knowledge in the feature extraction process.
In the end, we obtain tensors of size $[1, 1024]$ for both the video and audio as their respective features.

\subsection{Self-supervised Pre-training}
We perform self-supervised pre-training on the extracted audio features and video features. InfoNCE loss in the self-supervised training can be used to represent the distance (similarity) between two modalities and can be learned by training the InfoNCE loss to be smaller for the purpose of bringing the audio/video distance closer to the two correlations.

\hspace*{2em}For a batch size of $N$ data, employing InfoNCE loss , the model pairs the $N$ video features with the $N$ audio features, generating predictions for the similarity of $N^2$ possible video-audio pairs. The similarity here is calculated using the cosine similarity as described earlier. Among these, there are $N$ positive samples, corresponding to the true pairs of short videos and background music (diagonal elements in the matrix). The remaining $N^2 - N$ video-audio pairs serve as negative samples. The training objective aims to maximize the similarity among the $N$ positive samples while minimizing the similarity among the $N^2 - N$ negative samples.

\hspace*{2em}Specifically, for the 1024-dimensional video feature $\mathbf{v_{feat}}$, it is dimensionally reduced to 512 dimensions through an MLP linear layer, which is then passed through a one-dimensional batch normalization layer and a ReLU non-linear layer. Following these, a dropout layer with probability $p = 0.5$  and another MLP linear layer further reduce the video feature to 256 dimensions.
The audio feature undergoes the same processing steps as described above.
After these steps, the audio and video features are aligned in dimension. Regarding the feature representations of different modalities, we can consider them linked in a high-dimensional feature space. Here we use cosine similarity as the distance metric to effectively supervise the similarity between the features of the two modalities.

For the obtained 1024-dimensional audio feature $\mathbf{a_{feat}}$ and video feature $\mathbf{v_{feat}}$, the cosine similarity $\mathcal{L}_{\text{dist}}$ can be calculated as:

\[
\mathcal{L}_{\text{dist}} = \frac{\sum_{i=1}^{n} \mathbf{a}_{\text{feat}_i} \cdot \mathbf{v}_{\text{feat}_i}}{\sqrt{\sum_{i=1}^{n} (\mathbf{a}_{\text{feat}_i})^2 \cdot \sum_{i=1}^{n} (\mathbf{v}_{\text{feat}_i})^2}}
\]

where $\cdot$ denotes the dot product,n=1024.

\hspace*{2em}The final  $\mathcal{L}_{\text{NCE}}$ is formulated as follows:

\[
\mathcal{L}_{\text{NCE}} = -\sum_{i=1}^{bsz} \sum_{j=1}^{256} \log \frac{\exp (\mathbf{h}_s(\mathbf{y}_{i,j}^v, \mathbf{y}_{i,j}^a) / \tau)}{\sum_{i=1}^{bsz} \sum_{j=1}^{256} \exp (\mathbf{h}_s(\mathbf{y}_{i,j}^v, \mathbf{y}_{i,j}^a) / \tau)},
\]

where $\text{bsz}$ denotes batch size, 256 represents the dimensions of the embeddings for both modalities,
$h_s$ denotes the similarity function between the embeddings $y_i^v$ and $y_i^a$ of the $i$-th pair, $\tau$ is a temperature scalar, and the indices $i$ and $j$ represent the pairs and their components, respectively.

\begin{table}[t]
  \centering
  \caption{Comparison of different methods in terms of Recall@1. The best performance is highlighted in \textcolor{red}{red}, and the second best performance is highlighted in \textcolor{blue}{blue}.}\label{tab1}
  \setlength{\tabcolsep}{6pt}

  \begin{tabular}{l|ccccc}
    \hline
     Video \textbackslash Audio & VGGish & wav2vec2~\cite{baevski2020wav2vec} & YMNet & Musicnn~\cite{pons2019musicnn} & MVbind(Ours) \\
    \hline
    MobileNet~\cite{sandler2018mobilenetv2} & 0.4 & 0.0 & 0.0 & 0.2 & 0.0 \\
    CoAtNet~\cite{dai2021coatnet} & 0.2 & 0.0 & 0.0 & 0.2 & 0.0 \\
    MaxViT~\cite{tu2022maxvit} & 0.2 & 0.0 & 0.2 & 0.0 & 0.4 \\
    ResNet~\cite{he2016deep} & 0.0 & 0.2 & 0.0 & 0.2 & 0.6 \\
    Swin~\cite{liu2021swin} & 0.2 & 0.0 & 2.4 & 3.6 & 7.2 \\
    CLIP~\cite{radford2021learning} & 0.6 & 0.2 & 3.6 & 5.6 & \textcolor{blue}{9.2} \\
    MVbind(Ours) & 0.2 & 0.4 & 4.8 & 6.2 & \textcolor{red}{11.6} \\
    \hline
  \end{tabular}
\end{table}

\begin{table}[h]
  \centering
  \caption{Comparison of different methods in terms of Recall@5. The best performance is highlighted in \textcolor{red}{red}, and the second best performance is highlighted in \textcolor{blue}{blue}.}\label{tab2}
    \setlength{\tabcolsep}{6pt}
  \begin{tabular}{l|ccccc}
    \hline
     Video \textbackslash Audio & VGGish & wav2vec2~\cite{baevski2020wav2vec} & YMNet & Musicnn~\cite{pons2019musicnn} & MVBind(Ours) \\
    \hline
    MobileNet~\cite{sandler2018mobilenetv2} & 1.2 & 1.0 & 0.8 & 1.0 & 0.4 \\
    CoAtNet~\cite{dai2021coatnet} & 1.0 & 0.6 & 0.6 & 0.8 & 0.4 \\
    MaxViT~\cite{tu2022maxvit} & 1.0 & 1.2 & 0.6 & 0.6 & 1.0 \\
    ResNet~\cite{he2016deep} & 1.0 & 1.6 & 1.2 & 0.8 & 1.4 \\
    Swin~\cite{liu2021swin} & 1.6 & 0.8 & 10.4 & 10.8 & 21.2 \\
    CLIP~\cite{radford2021learning} & 1.6 & 0.8 & 17.4 & 17.6 & \textcolor{blue}{28.8} \\
    MVBind(Ours) & 1.4 & 1.2 & 12.6 & 16.2 & \textcolor{red}{33.4} \\
    \hline
  \end{tabular}
\end{table}

\section{Experiments}

We randomly selected 500 of the 8440 data in the dataset as the validation set and the remaining 7940 as the training set and tested the validation set matching results using the pre-trained model trained in the training set. We use the following CPU configuration: 4-core Intel(R) Xeon(R) CPU @ 2.30GHz. the GPU configuration is as follows: Nvidia Tesla P100.

\hspace*{2em}For the cross-modal retrieval task involving audio and video, the evaluation landscape lacks intuitive benchmarks and often heavily relies on subjective user assessments. One widely employed and effective evaluation metric is Recall@K, which measures the percentage of queries within the test set where at least one correct ground truth match is ranked among the top K matches. In this specific context, MVBind is designed to improve the Recall@K metric.

\hspace*{2em}Our validation process involves the assessment of 500 short videos accompanied by background music in our test set. To elaborate, each of these 500 samples is treated as a positive match, where the video and its corresponding background music are considered to be a correct pairing. Conversely, pairings with other samples are deemed incorrect matches. This rigorous validation approach serves to appraise the efficacy of MVBind in accurately retrieving relevant content in a cross-modal retrieval scenario.

\hspace*{2em}Currently, there exists a notable paucity of open-source initiatives focused on the cross-modal retrieval of audio/video tasks. The experiments conducted in this paper predominantly involve the comparative analysis of features derived from well-established audio/video processing models within each modality. These features are combined into 1024-dimensional tensors, and then self-supervised training is performed on these composite tensors. Subsequent to extensive experimentation, we decide to present the findings utilizing the Recall@1 (Table 1),  Recall@5 (Table 2), and Recall@10 (Table 3) metrics.
These findings provide valuable insights for future cross-modal studies, facilitating a more comprehensive understanding of which models are better at capturing relationships between different modalities.

\begin{table}[t]
  \centering
  \caption{Comparison of different methods in terms of Recall@10. The best performance is highlighted in \textcolor{red}{red}, and the second best performance is highlighted in \textcolor{blue}{blue}.}\label{tab3}
    \setlength{\tabcolsep}{6pt}
  \begin{tabular}{l|ccccc}
    \hline
     Video \textbackslash Audio & VGGish & wav2vec2~\cite{baevski2020wav2vec} & YMNet & Musicnn~\cite{pons2019musicnn} & MVBind(Ours) \\
    \hline
    MobileNet~\cite{sandler2018mobilenetv2} & 1.8 & 1.4 & 1.4 & 2.6 & 1.4 \\
    CoAtNet~\cite{dai2021coatnet} & 1.8 & 1.0 & 1.4 & 1.2 & 1.6 \\
    MaxViT~\cite{tu2022maxvit} & 2.6 & 3.0 & 1.4 & 1.8 & 2.2 \\
    ResNet~\cite{he2016deep} & 2.2 & 2.8 & 3.0 & 2.2 & 2.6 \\
    Swin~\cite{liu2021swin} & 2.8 & 1.4 & 17.2 & 17.4 & 29.8 \\
    CLIP~\cite{radford2021learning} & 3.6 & 2.4 & 23.2 & 27.8 & \textcolor{blue}{40.6} \\
    MVBind(Ours) & 3.8 & 2.2 & 21.8 & 25.6 & \textcolor{red}{45.6}\\
    \hline
  \end{tabular}
\end{table}

\hspace*{2em}The foundational processing networks employed in this paper encompass visual modalities such as CLIP~\cite{radford2021learning}, CoAtNet~\cite{dai2021coatnet}, MaxViT~\cite{tu2022maxvit}, MobileNet-V2~\cite{sandler2018mobilenetv2}, ResNet~\cite{he2016deep}, Swin Transformer~\cite{liu2021swin}, alongside auditory modalities including Musicnn~\cite{pons2019musicnn}, VGGish, wav2vec2.0~\cite{baevski2020wav2vec}, YMNet.(The abbreviation ``Swin Transformer" is shortened to ``Swin".The numbers in the table are all in percentage.)

\hspace*{2em}As illustrated in Table 1\&2\&3, The findings unequivocally demonstrate the underperformance of visual features extracted by networks such as CoAtNet, MaxViT, MobileNet, and ResNet in the realm of cross-modal retrieval. In contrast, CLIP and Swin Transformer exhibit superior performance. Both CLIP and Swin Transformer are specifically designed to jointly acquire knowledge from both text and image modalities. CLIP utilizes a contrastive learning method to train on images and text simultaneously, which enables the model to capture the intricate relationships between the two modalities. Moreover, the robust attention mechanism enables the model to establish meaningful connections across different components of the input. Conversely, networks like CoAtNet, MaxViT, MobileNet, and ResNet may be more tailored for image classification tasks. As a result,  the features extracted from these networks are better suited for single-modality
applications.

\begin{table}
  \centering
  \caption{Comparison with CBVMR.}
  \label{tab:tab4}
      \setlength{\tabcolsep}{10pt}
  \begin{tabular}{l|ccc}
    \hline
    & Recall@1 & Recall@5 & Recall@10 \\
    \hline
    CBVMR~\cite{hong2018cbvmr} & 0.2 & 1.0 & 1.8 \\
    MVBind(Ours) & 11.6 & 33.4 & 45.6 \\
    \hline
  \end{tabular}
\end{table}

\hspace*{2em}We also compared the CBVMR~\cite{hong2018cbvmr} with related studies using the method for experimental validation on our SVM-10K dataset, also using Recall and the results are shown in Table 4. The results show that there is a huge difference in the effectiveness of the two methods because the CBVMR method is not able to effectively learn the correlation between the two methods when dealing with the broad-modal retrieval task, which is a more complex task, and the datasets used by the two methods are very different, which results in the trained network not being applicable to the SVM-10K dataset.

\hspace*{2em}The proposed MVBind has a distinct advantage over alternative combinations, primarily due to the utilization of the pre-trained ImageBind model in the feature extraction network. The ImageBind model has effectively aligned the features of the audio and video modalities, providing a conducive environment for subsequent self-supervised learning networks to glean higher-level associations, like emotions.

\section{Conclusion}
\label{sec:conclusion}

In summary, this paper introduces MVBind, a self-supervised music-video embedding space-binding model towards cross-modal retrieval for short videos.
We first establish a short video with music-10K database, termed SVM-10K. Filling the gap in automatic music recommendation, our model, trained on the SVM-10K dataset, outperforms baseline methods. By addressing the complexities of short video dynamics, such as emotions and aesthetics, MVBind offers an efficient solution for music-video alignment, promising improved content creation experiences in the era of short videos.

%
%
%
\bibliographystyle{splncs04}  
\bibliography{reference} 

\end{document}